\documentstyle[psfig]{l-aa}
\begin{document}
   \thesaurus{12         
              (11.03.4 Abell 370;  
               11.05.2;  
               12.03.3;  
               12.07.1)} 

\title{Lensed galaxies in Abell 370}
\subtitle{I. Modeling the number counts and redshift distribution of 
background sources
\thanks{Based on observations with the NASA/ESA {\it Hubble 
Space Telescope\/} obtained from the data archive at the Space Telescope 
European Coordinating Facility and with the 3.6m Telescope of the 
European Southern Observatory, La Silla, Chile and the Canada-France-Hawaii
Telescope at Mauna Kea, Hawaii, USA.}
}

   \author{
J. B\'ezecourt\inst{1, 2} \and 
J.P. Kneib\inst{1} \and
G. Soucail\inst{1} \and
T.M.D. Ebbels\inst{3} 
   }

   \offprints{J. B\'ezecourt, bezecour@astro.rug.nl}

   \institute{Observatoire Midi-Pyr\'en\'ees, Laboratoire d'Astrophysique, 
    UMR 5572, 14 Avenue E. Belin, F-31400 Toulouse, France 
   \and Kapteyn Institute, Postbus 800, 9700 AV Groningen, The Netherlands
   \and Institute of Astronomy, Madingley Road, Cambridge CB3 0HA, UK}

\date{Received , Accepted }

\maketitle


\begin{abstract} 
We present new observations of the cluster-lens Abell 370: a deep
HST/WFPC2 F675W image and ESO 3.6m spectroscopy of faint galaxies.
These observations shed new light on the statistical properties of
faint lensed galaxies.  In particular, we spectroscopically confirm
the multiple image nature of the B2--B3 gravitational pair
(\cite{kneib93}), and determine a redshift of $z=0.806\pm 0.002$ which
is in very good agreement with earlier predictions.  A refined mass
model of the cluster core (that includes cluster galaxy halos) is
presented, based on a number of newly identified multiple
images. Following B\'ezecourt et al. (1998a), we combine the new cluster
mass
model with a spectrophotometric prescription for galaxy evolution to
predict the arclets number counts and redshift distribution in the HST
image.  In particular, the ellipticity distribution of background
sources is taken into account, in order to properly estimate the
statistical number and redshift distribution of arclets.  We show that
the redshift distribution of arclets, and particularly its
high redshift tail can be used as a strong constraint to disentangle
different galaxy evolution scenarios. A hierarchical model which
includes number density evolution is favored by our
analysis. Finally, we compute the depletion curves for the faint
galaxy number counts and discuss their wavelength dependence.

\end{abstract}

\section{Introduction}
Thanks to gravitational amplification, galaxy clusters are a powerful
tool to probe distant galaxies.  Indeed, one of the most distant
galaxy detected is a gravitational arc at $z$=4.92 in the cluster
Cl1358+62 (\cite{franx97}).  Other very distant lensed sources ($z>$3)
have been identified behind Cl0939+4713 (\cite{trager97}) and A2390
(\cite{pello98a}, \cite{frye98}) leading to important results on
the
formation history and evolution of galaxies at large redshift.
Determining the redshift of arcs and arclets is of great importance as
it fixes the angular scales of the optical configuration, hence giving
an absolute cluster mass estimate within the arc radius ({\it e.g.}
\cite{soucail88}, \cite{mellier91}).  But despite the cluster
magnification, measuring arclets redshifts is a difficult
observational task due to their low surface brightness
(\cite{bezecourt97}, \cite{ebbels98b}), and the lack of strong
spectral features in the optical domain for galaxies with
$1<z<2.5$.

Accurate modeling of cluster potentials based on the analysis of
multiple images and weak shear distortions has shown that cluster
mass distributions are best described by the sum of a smooth and large
scale component (the cluster) and the contribution of cluster galaxy
halos (\cite{kneib96}, \cite{natarajan98}, \cite{geiger98}). For a given
mass distribution, Kneib et al (1996) demonstrated that a redshift
can be estimated if one can measure accurately the shape of an
individual arclet. In order to validate this method and study its
biases, extensive programs of gravitational arclet spectroscopy
have
been undertaken. In particular, Ebbels et al. (1998) have measured 19
redshifts of lensed objects behind Abell 2218. Most of them confirm
the lens redshift prediction, and allow the accuracy of the mass
model to be improved. Similarly, B\'ezecourt and Soucail (1997) have
started the spectroscopy of arclets in Abell 2390, which has been used
to constrain the mass distribution in this cluster with a great
accuracy (\cite{kneib98}).

Using these accurate cluster mass models and a spectrophotometric
description of galaxy evolution (Bruzual \& Charlot 1993,
\cite{pozzetti96}), 
B\'ezecourt et
al. (1998a) have predicted the expected arclet
number counts and their redshift distribution.  This model presents
many improvements with respect to previous work ({\sl e.g.}
\cite{nemiroff89}, \cite{wu93}, \cite{grossman94}, \cite{hattori97}) 
as it includes many observational limits such as magnitude ranges,
surface brightness cut-off or a choice of the optical waveband, and
this for any mass distribution, regardless of its complexity.

Abell 2218 is the first cluster where the number counts and redshift
distribution of the background arclets have been examined in 
detail
(\cite{bezecourt98}, \cite{ebbels98b}). We propose in this paper to
further extend this study to another well known cluster lens, namely
Abell 370.  Its mass distribution was first accurately derived by
Kneib et al. (1993) [hereafter K93] who showed that the mass model has
to be bimodal in shape in order to accommodate the gravitational pair
B2--B3. This was later confirmed by the HST/WFPC1 observation described
in Smail et al. (1996) and the X-ray map of the cluster, displaying a
bimodal shape compatible with the lens model (\cite{fort94}).

We present new observations of the cluster Abell 370 in Section 2: a
deep HST/WFPC2 image and spectroscopic data. Section 3 discusses the
new lensing mass model. In section 4, we use an improved version of
the code developed by B\'ezecourt et al. (1998a) to determine the
expected counts and redshift distribution of arclets in Abell 370.
Our analysis explores two different scenarios of galaxy evolution
to study their differences, and compute the depletion curves of
background number counts at different wavebands. Discussion,
conclusions and further prospects to constrain galaxy evolution
through
lenses are presented in Section 5.  Throughout the paper, we use
a Hubble constant of H$_0 = 50 \, {\rm km \,s}^{-1} {\rm Mpc}^{-1}$,
with $\Omega_0$= 0 or 1 and $\Omega_{\Lambda} = 0$.

\section{New observational data}
\subsection{HST observations and photometry}
Abell 370 was observed with the HST/WFPC2 camera with the
F675W filter [ID: 6003, P.I.:R.P. Saglia], resulting in a reasonably
deep image with a total exposure time of $T_{exp}$ = 5.6 ksec.  These
data were provided by the ESO/ST-ECF Science Archive Facility,
Garching, Germany and were reduced with standard reduction
procedures. We used the IRAF/STSDAS packages for the image
combinations after centering and cosmic ray cleaning.  The
absolute photometry was obtained using magnitude zero-points given in
Holtzman et al. (1995). The photometry of the field was obtained with
the Sextractor package (\cite{bertin96}). A criterion of 12
contiguous
pixels above the given detection threshold was used to identify an
object.  The 1 $\sigma$ detection limit is $R_{675W}=24.9$ mag
arcsec$^{-2}$. From the magnitude histogram obtained from the catalog,
we estimate the completeness limit to be $R_{675W}=25.5$.

We also built a sample of arclets for the purpose of our study of
their photometric and statistical properties.  To define an arclet we
imposed the following selection criteria: at least 12 contiguous
pixels above 2$\sigma$ of the local sky level, an axis ratio greater
than 2, a central surface brightness lower than $\mu_{675W}=25.5$
mag. arcsec$^{-2}$ and a magnitude range $20<R_{675W}<26$.  The final
sample contains 81 arclets and their magnitude histogram is discussed
in Section 4.2.

\subsection{Identification of multiple images and arclets}
\begin{figure*}
\centerline{\psfig{figure=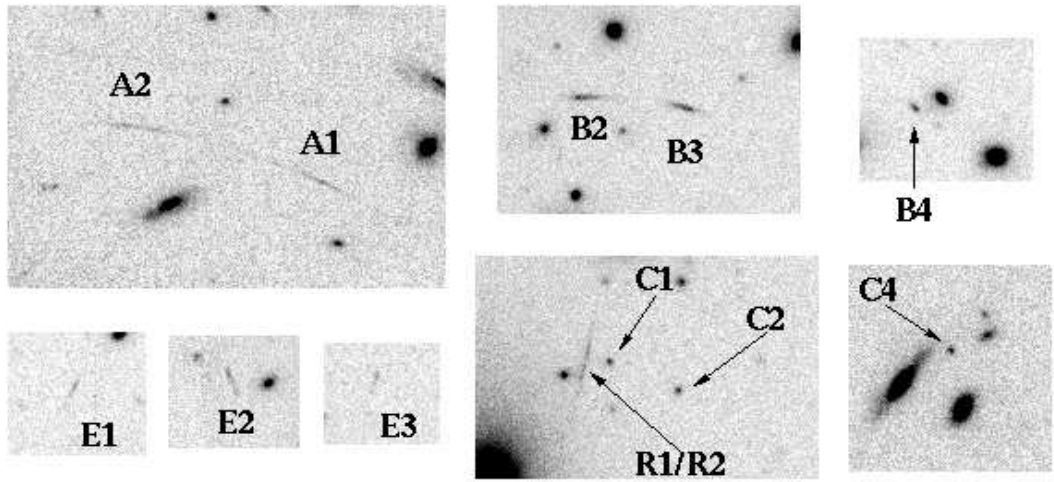,width=0.8\textwidth}}
\caption{Detailed view of the multiple image candidates detected
in the
WFPC2/F675W image. B2--B3--B4 is a triple image configuration, as well
as C1--C2--C4. R1/R2 is a radial arc.
E1--E2--E3 is also a triple configuration with a clear inversion
of parity between E2 and E1/E3 (see text for more details).  }

\label{ima_arclets}
\end{figure*}

Abell 370 ($z$=0.37) is a rich optical cluster dominated by two giant
elliptical galaxies identified as \#20 and \#35, following the
numbering of Mellier et al. (1988). A set of multiple image
candidates and gravitational arcs are identified on the WFPC2/F675W
image and are displayed in Figure \ref{ima_arclets}. Their photometric
and geometrical properties are also summarized in Table
\ref{table_arclets}. We now discuss them in detail:

\noindent{\bf A0 : } Near galaxy \#35
is located the spectacular giant arc ($z$=0.724) initially detected by
Soucail et al. (1987). From the high resolution WFPC2 image more details
on its morphology are clearly visible, suggesting that it is a
gravitationally lensed image of a spiral galaxy (\cite{soucail98}).

\noindent{\bf A1 to A6 : }
A first set of arclets (labelled A1 to A6) was detected from
ground based images by Fort et al.  (1988) and further discussed by
Kneib et al. (1994). Most of them are blue objects, but none have a
spectroscopic redshift yet. A5 is the most extended one and presents
very blue colors and a strong dimming in the reddest bands, suggesting
a young star forming galaxy. Despite deep spectroscopic data
for A5 (\cite{mellier91}) no significant
emission line has been identified suggesting that $1<z_{A5}<2.2$.
Arclets A3 to A6 are probably single images of background lensed
galaxies in view of their location in the lens plane. Arclets A1 and
A2 may be multiple images of the same source.
 
\noindent{\bf The multiple images B, C and D : }
K93 demonstrated that the B2--B3 objects (Fort et al. 1988) correspond
to a gravitational pair with a counter image labeled as B4.  The
B2--B3 pair and the arc A0 were used to constrain the K93 model which
showed a bimodal mass distribution.  This model proposed that objects
C1--C2--C3 and D1--D2, identified on high quality ground-based images,
were also multiple image systems. A redshift estimate based on the
mass model was proposed for each of the pairs: $z_B= 0.865$, $z_C=
0.81$ and $z_D= 0.95$.  The reality of B4 was confirmed {\it a
posteriori} by the HST/WFPC1 data (Smail et al. 1996).  C3 is likely
to be a wrong identification of the counter image of C1--C2, and we
denote C4 the correct counter image used in the present model.

\noindent{\bf The radial arc R : }
In the HST/WFPC1 image, Smail et al. (1996) discovered a radial arc
candidate R (R1-R2) located close to galaxy \#35. They modeled it as a
5-image configuration, and predicted a redshift of $z\sim 1.3$ using
the K93 model.  
This radial arc is well identified in the F675W image and is clearly
the merging of 2 images.

\noindent{\bf New multiple images : }
The detailed insight of the mass model and the exquisite HST
resolution allow the identification of other multiple image
candidates. We only discuss here the E1--E2--E3 multiple configuration
as it presents a characteristic inversion of parity as expected from
lensing theory (see also \cite{smail95} and
\cite{colley96} for other spectacular examples). Other multiple 
image candidates will be discussed in a forthcoming paper.

\begin{table}
\caption[]{Main geometrical and photometric properties of the multiple 
images candidates identified on the F675W HST image. The origin of the
coordinates is the center of galaxy \#35 and the XY orientation is the
CCD one, {\em i.e.} North is to the top with an angle of 85$^\circ$
41$'$ clockwise from the X axis and East is to the left. Coordinates
are in arcseconds.  $e$ is the ellipticity of the objects ({\em i.e.}
1 -- b/a where b/a is the axis ratio of the isophotes) and $\theta$ is
the orientation of the isophote with respect to the X axis. Surface
photometry was computed on the two brightest elliptical galaxy using
the {\it ellipse} package in STSDAS.  The two galaxies are well fitted
by a de Vaucouleurs law with $R_e= 32.6 h_{50}^{-1}$ kpc for \#20 
and $R_e=44 h_{50}^{-1}$ kpc for \#35.
 }
\label{table_arclets}
\begin{flushleft}
\begin{tabular}{crrcrc}
\hline\noalign{\smallskip}
Object & X ($''$) & Y ($''$) & $e$ & $\theta \quad$ & R$_{675W}$ \\
\noalign{\smallskip}
\hline\noalign{\smallskip}
A1 &   16.35 & 57.30 & 0.74 & -23.8 &  24.83 \\
A2 &    5.56 & 60.43 & 0.75 &  -4.6 &  24.38 \\
B2 &    8.46 & 21.14 & 0.76 &  -0.2 &  23.08 \\
B3 &   13.80 & 20.58 & 0.74 & -15.7 &  23.23 \\
B4 &  -19.45 & 20.98 & 0.42 & -41.6 &  23.82 \\
C1 &    6.46 &  6.50 & 0.17 &   4.6 &  24.05 \\
C2 &   10.19 &  4.96 & 0.10 &  63.2 &  24.00 \\
C3 &  -21.82 &  2.78 & 0.14 & -60.9 &  23.99 \\
R1/R2 & 5.27 &  7.03 & 0.76 &  76.4 &  23.98 \\
E1 &   32.86 & 18.62 & 0.51 &  74.6 &  24.83 \\
E2 &  -31.22 & 19.15 & 0.68 &  64.4 &  24.53 \\
E3 &    0.54 & 22.21 & 0.66 & -69.8 &  24.44 \\
\noalign{\smallskip}
\hline
\noalign{\smallskip}
\# 20 & 3.09 & 37.85 & 0.36 & -70$\pm 3$ & 16.9 \\
\# 35 & 0.00 & 0.00 &  0.45 & -77$\pm 4$ & 17.1\\
\noalign{\smallskip}
\hline
\end{tabular}
\end{flushleft}
\end{table}

\subsection{Spectroscopic Observations}
Spectroscopic data were acquired at the 3.6m telescope of La Silla
(ESO) with EFOSC on October 1996. A long slit of 1.5\arcsec\ width was
positioned along the two objects B2 and B3 for a total integration
time of 3 hours with an average seeing of 1.9\arcsec. Following the
predictions of K93, the [O{\sc ii}] emission line was expected to lie
at $\lambda\sim 6950$\AA .  Thus the R300 grism was used, providing a
useful spectral range 6000\AA --9000\AA\ and a dispersion of
7.5\AA/pixel.  The data were reduced with standard procedures for
flat fielding, wavelength and flux calibration in the IRAF
environment. Sky subtraction was performed on the 2D image, and the
final spectra were extracted with an optimal extraction algorithm.

\begin{figure}
\psfig{figure=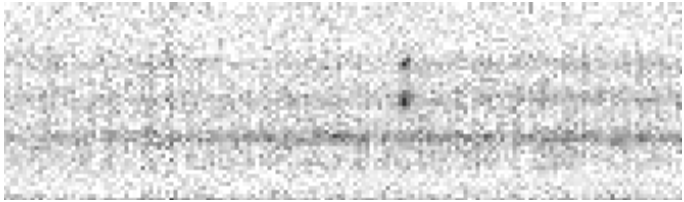,width=0.5\textwidth}
\caption{Two dimensional spectrum of objects B2 and B3. 
The [O{\sc ii}] emission line is visible for the 
spectrum of B2 (bottom) and B3 (top) in the center of the image.}
\label{fig-B2B3}
\end{figure}

\begin{figure}
\psfig{figure=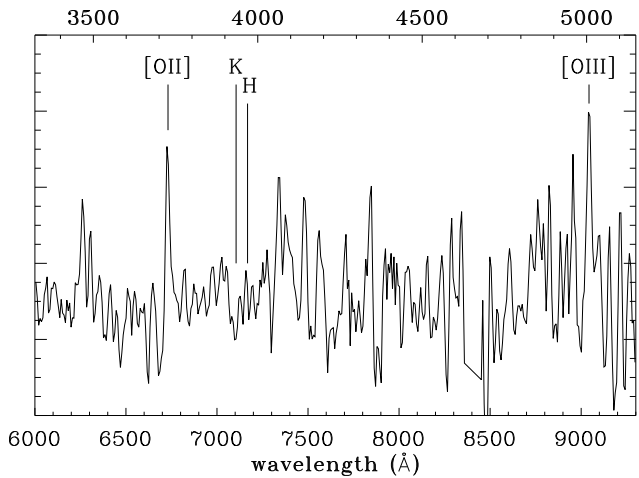,width=0.5\textwidth}
\caption{Spectra of objects B2 and B3 co-added and flux calibrated. 
The ordinate is $F_{\lambda}$ in arbitrary units with zero level
at the bottom of 
the graph. The wavelength scale is the observed one at the bottom, and 
the rest frame one at the top. The two emission lines detected in the
spectra 
are quoted with their identification, as well 
as the location of the absorption H and K lines, not clearly detected.
}
\label{fig-specB2B3}
\end{figure}

Both objects B2 and B3 show a prominent emission line at
$\lambda=6728$\AA\ and $\lambda=6727$\AA\ respectively on the two
dimensional spectra, which we identify with [O{\sc ii}] 3727\AA\
(Figure \ref{fig-B2B3} and \ref{fig-specB2B3}).  Another feature
appears at $\lambda=9045$\AA\ and $\lambda=9042$\AA\ respectively,
identified with [O{\sc iii}] 5007\AA\ thus confirming the
redshifts. We find $z_{B2}=0.8058$ and $z_{B3}=0.8054$, clearly
demonstrating the similarity of the two objects within the error bars,
and in good agreement with the $z=0.865$ prediction. This definitely
confirms the nature of these objects as a multiply imaged galaxy.

Although, arclets A1 and A2 were observed during 4.5 hours, no
continuum nor emission line could be detected.

\subsection{A closer look at the  B2--B3--B4 multiple image}
%
\begin{figure}
\psfig{figure=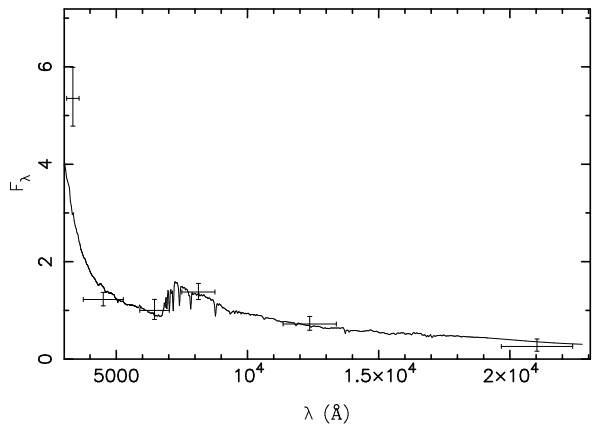,width=0.5\textwidth,angle=-90}
\caption{Multicolor photometry of object B3 in $U, B, R, I, J$ and $K'$.
The best fit of the data points corresponds to a star forming galaxy with a
constant star formation rate seen at $t=1.7$ Gyr, with $Z=Z_{\odot}$ 
and a null absorption. Magnitudes are: 
$U = 22.43 \pm 0.1$, $B = 24.58 \pm 0.1$, $R = 23.84 \pm 0.2$,
$J = 22.08 \pm 0.2$, $K' = 21.26 \pm 0.5$.
}
\label{fig-SED-B3}
\end{figure}

This multiple image is one of the bluest arclets observed in this 
field. Accurate photometry is available in many filters extending from 
$U$ to $K'$ with data coming from a U HST image with the F336W filter 
(ID: 5709, P.I. J.M Deharveng, \cite{paper2}, hereafter Paper II), $B$, $R$ 
and $I$ images from CFHT 
(\cite{kneib94}) and unpublished  CFHT infrared images in $J$ and $K'$ bands
(Le Borgne and Picat, private communication). 
B2 lies too close to galaxy \#20 and is contaminated by its envelope at
long wavelengths, hence we only concentrate on B3. 

We compute the spectral energy distribution (SED) of B3, corrected
from its redshift, in order to study the stellar content of the
galaxy. We compare and fit the B3 SED with synthetic ones from the
Bruzual
and Charlot spectrophotometric evolutionary code (GISSEL96).  This
code can provide SEDs for different metallicities ranging from
Z$_{\odot}$/50 to 5 Z$_{\odot}$, different spectroscopic galaxy type
determined by the star formation history (burst, elliptical, spiral
and irregular), and different internal absorption by dust can be
studied. The extinction law used is the one proposed by Calzetti
(1997) which is similar to what is observed in our galaxy without the
2175\AA\ bump.  We explore the 3-dimensional parameter space
(spectroscopic type, metallicity and absorption, E($B-V$) from 0 to
0.3 mag.) to fit the B3 photometry by the model SEDs.

\begin{figure}
\psfig{figure=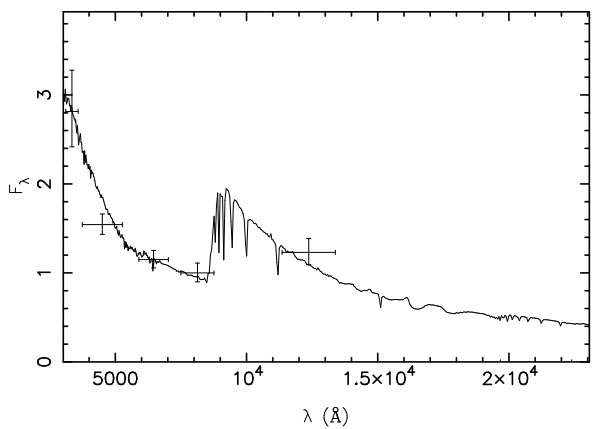,width=0.5\textwidth,angle=-90}
\caption{Multicolor photometry of object A5 in $U, B, R, I,$ and $J$
(undetected in $K'$).  The best fit of the data points corresponds to
a burst of star formation redshifted at $z=1.3$ 
and seen at $t=0.3$ Gyr, with $Z=Z_{\odot}/50$ and $E(B-V)=0.1$
mag. Magnitudes are: $U=21.70 \pm 0.15, B= 22.90 \pm 0.04, R=22.26\pm
0.06$, $I=21.67\pm 0.09, J= 20.07\pm 0.11$.}

\label{fig-SED-A5}
\end{figure}

The main constraint comes from the high UV flux which requires a 
weak internal absorption ($E(B-V)<0.2$mag) and a low metallicity
($Z<Z_{\odot}$) (Figure
\ref{fig-SED-B3}).  Despite this, the fit does not reproduce the UV
part very well. A possibility is that we are seeing the superposition of a
very
recent burst of star formation on an older stellar population. A
strong starburst alone would show a much fainter flux at red and
infrared wavelengths, hence the observed IR emission must come from an
already existing old population. The B multiple image is similar to
arclet A5 which is also very bright in the UV but much fainter in the
IR ($R-K'=2.6$ for B3 and $R-K'<1.9$ for A5). More precisely, assuming 
a redshift of
$z$=1.3 (Mellier et al. 1991), we have similarly fitted the A5 SED.
It shows evidences for a low absorption ($E(B-V)=0.1 $ mag.) and a
very low metallicity ($Z=Z_{\odot}/50$) (Figure \ref{fig-SED-A5}). 
The fact that the best-fit metallicity ($Z_{\odot}/50$) is
the lower bound allowed by the code should not be a problem as a satisfying 
fit is also obtained for $Z_{\odot}/5$ with comparable significance.
The trend towards low $Z$ is then stable and not too fast.
The corresponding age is 0.3 Gyr, drawing the portrait of a very young
object in an active phase of star formation. An additional old
population is not required for A5 which is the major difference with
object B.

These two objects B3 and A5 are typical examples of the population
revealed by UV imaging. The UV selection exhibits young objects with
low metallicity and low absorption in the redshift range
[0.5,2.0]. 

Such detailed stellar population study, could in principle be applied
on any other arclets in the field, for which redshifts have been
spectroscopically confirmed or are strongly constrained by lensing and 
photometric redshift techniques (Pell\'o et al. 1998a).  

\section{An improved lens model}
\begin{figure*}
\psfig{figure=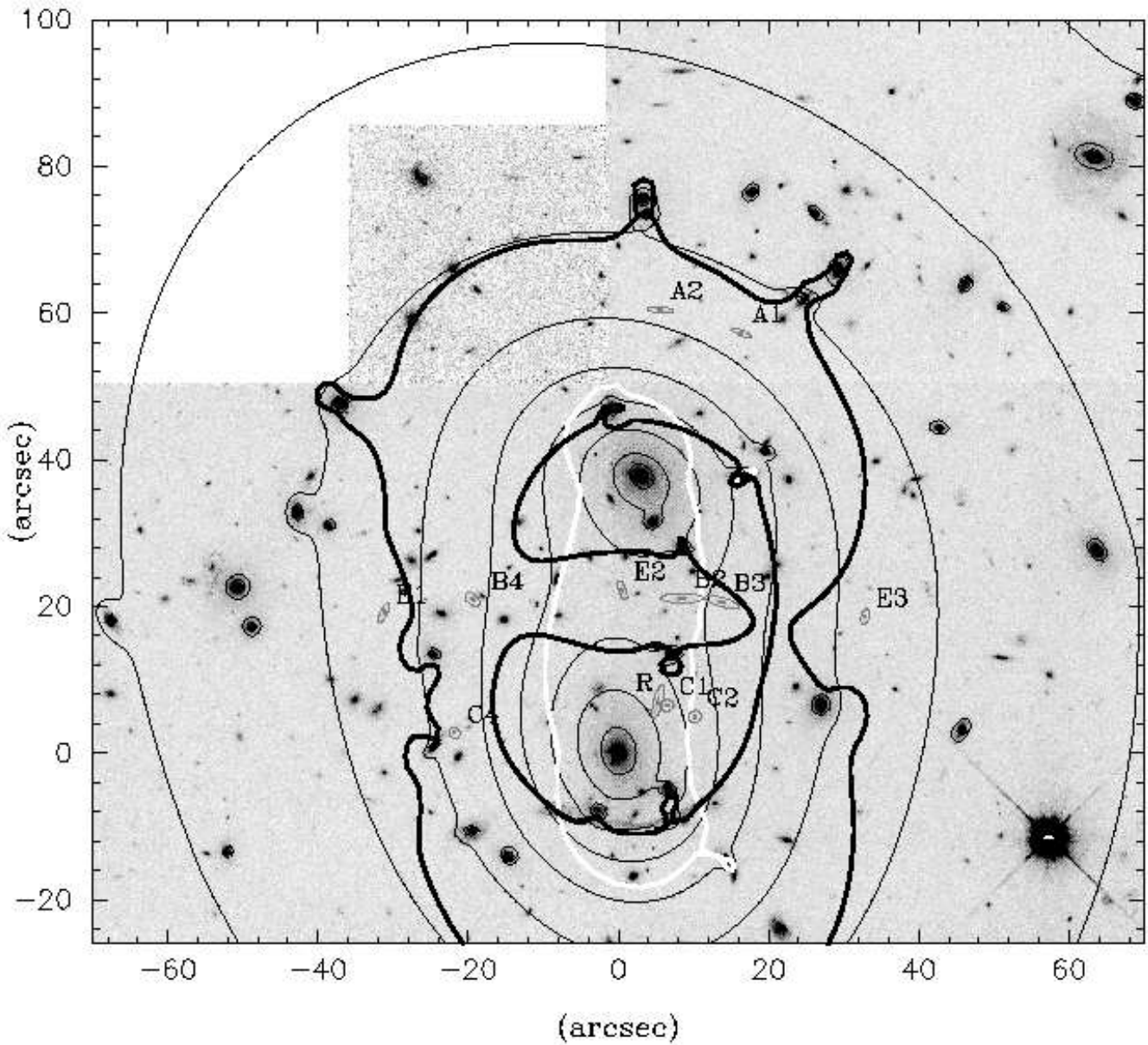,width=\textwidth}
\caption{Central part of Abell 370 seen with WFPC2 in F675W.
Overlaid is the mass distribution (black thin lines) and the critical
lines at $z=0.806$ (thick white lines) and $z=4$ (thick black
lines). The most important arclets and multiple images are shown.
}
\label{fig-modele}
\end{figure*}

Using both the giant arc A0 and the B2--B3--B4 triple system, K93
showed that a bimodal mass distribution was an adequate fit to these
multiple image constraints (see also AbdelSalam et al. 1998 for a
different 
approach). Here, we improve this gross picture by
taking into account the contribution of the cluster galaxies (mainly
E/S0 galaxies in the cluster core) in a similar way as Kneib et
al. (1996) and Natarajan et al. (1998). All mass components
(galaxies+clusters) have been assumed to follow a truncated
pseudo isothermal mass distribution (PIEMD, Kassiola and Kovner 1993,
Hjorth \& Kneib 1998).

Following Kneib et al. (1996), for each galaxy halo,
the velocity dispersion $\sigma_0$,
the truncation radius $r_t$ and the core radius $r_0$
are scaled to the galaxy luminosity computed from the
observed F675W magnitude.
The scaling relations used for the galaxy halos are: 
\begin{equation}
\sigma_0=\sigma_{0\ast} \left({L\over L_{\ast}}\right)^{{1\over 4}},
\end{equation}
\begin{equation}
r_t=r_{t\ast} \left({L\over L_{\ast}}\right)^{{0.8}},
\end{equation}
\begin{equation}
r_0=r_{0\ast} \left({L\over L_{\ast}}\right)^{{1\over 2}}.
\end{equation}
The scaling relations adopted are motivated by the properties of the 
fundamental plane (FP)
and are similar to the one used by Brainerd et al. (1996). In
particular, the exponent 0.8 in eq. (2) leads to a total mass-to-light
ratio that scales with $L^{0.3}$ in agreement with the observed
correlation of the FP ({\it e.g.} Jorgensen, Franx \& Kjaergaard
1996). Brighter galaxies have more extended and therefore more massive dark
halos.

The orientation and ellipticity of the galaxy halos are taken from the
observed values of the light distribution while $\sigma_{0\ast}$ and 
$r_{t\ast}$ are optimized. $r_{0\ast}$ is fixed at 0.15 $kpc$.
To model the cluster
components we considered two `large scale' mass distributions
(represented by 2 truncated PIEMD mass distribution) centered on 
galaxies \#20 and \#35. Their orientations, ellipticities, velocity 
dispersions and core radius are left as free parameters.

The constraints imposed in the optimization procedure are the multiple
images described in section 2.2, along with the redshifts of A0 and of
the B2--B3--B4 triplet.  The best fit parameters are summarized in Table
\ref{hst} ($\chi^2$=4.5).
We also derive a ``lensing'' redshift estimate for the multiple images.  The
C
system is at $z_C=0.75\pm 0.1$, D is at $z_D=0.85\pm 0.1$ and E is at
$z_E=1.3\pm 0.1$. R is moved to larger redshift with $z_R=1.7\pm 0.2$
and A1/A2 could form a gravitational pair at $z_{A1/A2} =1.4 \pm 0.2$.  
We computed the total projected mass in different apertures centered
at the barycentre of the \#20 and \#35 galaxies. 
We find that within
75, 150 and 300 kpc the total mass is, respectively, for this modeling 
and the previous one (K93):
0.5 (0.45) $\pm$0.05 10$^{14}$ M$_\odot$,
1.8 (1.6) $\pm$0.1 10$^{14}$ M$_\odot$ and
4.8 (4.3) $\pm$0.15 10$^{14}$ M$_\odot$.
Furthermore, 5\% of the total mass is retained in cluster galaxy
halos. The total mass-to-light ratio is $\sim$180 (M/L$_V$)$_\odot$ 
(out to 300kpc, 160(M/L$_V$)$_\odot$ for the K93 modeling), 
and $\sim$9 (M/L$_V$)$_\odot$ for a $L_{\ast}$
galaxy halos. These results are similar to the one found in other
cluster lenses like Abell 2218, 2390 and AC114 (Kneib et al 1996, 1998;
Natarajan et al 1998).

\begin{table}
\caption{Model parameters for A370 potential}
\label{hst}
\begin{flushleft}
\begin{tabular}{ccccccc}
\hline
id&$\varepsilon$&$\theta$&$\sigma_0$&$r_0$&$r_t$\\
 &${a^2-b^2\over a^2+b^2}$ & &$km\,s^{-1}$&$kpc$&$kpc$\\
\hline
35&0.23&-80&1050&75&800\\
20&0.12&-57&1100&91&800\\
galaxies& -- & -- &$\sigma_{0\ast}=125 $&$r_{0\ast}=0.15
$&$r_{t\ast}=15 $\\
\hline
\end{tabular}
\end{flushleft}
\end{table}

\section{Statistical properties of faint galaxies}

\subsection{Modeling the number counts of lensed galaxies}
The number counts of gravitational arclets in the field of a massive
cluster of galaxies are a competition between
the magnification of the luminosity by the cluster
potential that makes more objects visible and the surface
dilution that
decreases the surface density of arclets by the same factor as for the
magnification.

The number of arcs brighter than magnitude $m$ with an axis ratio
greater than $q_{min}$ and a surface brigthness brighter than $\mu_0$
within the field of a cluster of galaxies is:
\begin{equation}
\begin{tabular}{ccc}
$\lefteqn{N(m,q_{min},\mu_0) = }$ \\
 & & \\
 & & $ \sum_{i} \int_{z_l}^{z_{max}} \int_{q_{min}}^{\infty}
S(q,z) \ \int_{L_{min}}^{L_{max}} \Phi_i(L,z) \, dL \, dq
{dV \over dz} \, dz $
\end{tabular}
\end{equation}
The sum is over the different morphological types $i$.
$z_l$ is the lens redshift and $z_{max}(\mu_0,i)$ is the redshift cutoff
corresponding to the limit in central surface brightness $\mu_0$.
$S(q,z)$ is the angular area in the source plane
(at redshift $z$) that gives arcs with an axis ratio between $q$ and
$q+dq$. $\Phi_i(L,z)$ is the luminosity function at redshift $z$ for each 
morphological type.

It was required as
a preliminary step that counts in wavebands from $U$ to $K$ and redshift 
distributions in $B$ and $I$ in empty field are correctly reproduced.
We used the model for galaxy evolution of Bruzual and Charlot (1993)
with the prescriptions of Pozzetti et al. (1996) for the
optimisation of the parameters representing the field galaxy
distributions (see B\'ezecourt et al.  1998a for details).

The weak point in the B\'ezecourt et al. 1998a model, is the
approximation of circular sources used for the calculation of arclets
axis ratio. Indeed, this may underestimate the number of galaxies with
axis ratio larger than a certain threshold especially at low redshifts
and then bias the predicted redshift distribution.  To provide a more
accurate and reliable redshift distribution, we now take into account
the galaxy ellipticity distribution as given by Ebbels (1998) (this
distribution is based on the analysis of HST/MDS-like images, and is
therefore appropriate to our study):

\begin{equation}
p(\tau) \propto \tau \, \exp \left[- \left({\tau \over 0.036}\right)^{0.54} 
\right]
\label{ell}
\end{equation}

where $\tau={ a^2-b^2 \over 2\, a \, b }$ for an elliptical object
with semi major axis $a$ and semi minor axis $b$.
The complex deformation  $\vec \tau$ is defined by: $\vec \tau = \tau
\cos (2\theta)+ i\tau \sin (2\theta)$ where $\theta$ is the
orientation of the major axis.  At each point in the image
plane (and at each redshift step for the sources), we estimate the
fraction of the background galaxies which can give a lensed image more
distorted than a fixed limit.  We first consider the effect of the
lens deformation through the complex relation given by
Kneib et al. (1996) that relates the image deformation
$\vec \tau_I$ as a function of the
source deformation $\vec \tau_S$ and the strength of the potential
$\vec \tau_{pot}$:

\begin{equation}
{\rm sgn}(\det{A^{-1}}){\vec \tau_I}=
{\vec \tau_S}+{\vec \tau_{pot}}
\left(\delta_S + \Re( \vec \tau_S\vec g^*_{pot})\right)\ 
\label{eq:lenscomplex}
\end{equation}
where
 $\delta  = \sqrt{\left( 1 + \tau^2 \right)} = 1 + \vec g \vec \tau^*$
(see Kneib et al 1996 for a full description of this formalism).
The term $\Re( \vec \tau_S\vec g^*_{pot})$ is a correction for 
strong lensing only.
A lower threshold in the observed axis ratio ($a/b>2$ here) 
corresponds to a lower threshold in the deformation ($\tau_I > 0.75$)
whatever the position angle of the image.  Within this limit, we then
scan all the allowed solutions for $\vec\tau_S$ and the source
position angle $\theta_S$. The fraction ${\cal F} (x_I,y_I,z)$ of galaxies 
at the location $(x_I,y_I)$ in the image plane and at redshift $z$ which
fill
the condition is computed using the probability distribution of equation
(4) for $\vec\tau_S$ and an average over the position angle
$\theta_S$.


\begin{equation}
{\cal F} (x_I,y_I,z) =\frac{1}{2 \pi} \int_{0}^{2 \pi} \int_{\vec\tau_S\ 
|\  \tau_I>0.75} p(\tau_S)d\tau_S d\theta_S
\end{equation}

\subsection{Comparison between the observed and predicted arclets 
number counts}

\begin{figure}
\psfig{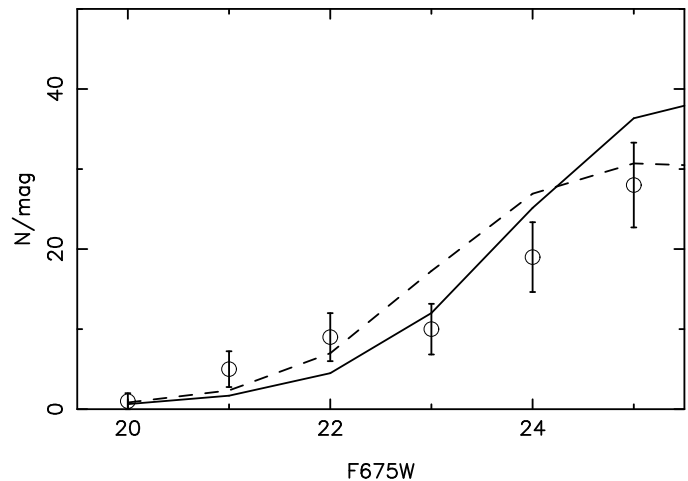}
\caption{Number counts of arclets in A370 in the F675W HST image with the 
following selection criteria: $a/b>2$ and $\mu_{675W}<25.5$. Observed
counts ($\circ$) are compared to two number counts models with
respectively $q_0=0.5$ (solid line) and $q_0=0$ (dashed line).  Error
bars are statistical poissonian uncertainties. }
\label{histo_F675W}
\end{figure}

The magnitude histogram of the arclets observed in A370 is shown in
Figure \ref{histo_F675W}. As already stressed in B\'ezecourt et
al. (1998a), we find that the addition of galaxy scale components and
the effect of the source ellipticity distribution significantly
increase arclets counts by a factor of 1.4 ($q_0=0.5$) and 1.7
($q_0=0$) with respect to a K93 model with the assumption of circular
sources.  Such increase in the arclets counts was already observed in
B\'ezecourt et al. (1998a) when using a detailed mass distribution 
but only with circular sources. 
Considering an ellipticity distribution for the galaxies also
increases the arclet counts and strongly modifies the redshift distribution 
as discussed below. With this refinement, the arclet counts
prediction is now more consistent with the data over the whole
magnitude range.  Clearly, detailed lens models and proper assumptions
on the source ellipticity distribution are mandatory to explore
arclets counts.

\subsection{Redshift distribution of arclets}
\begin{figure}
\psfig{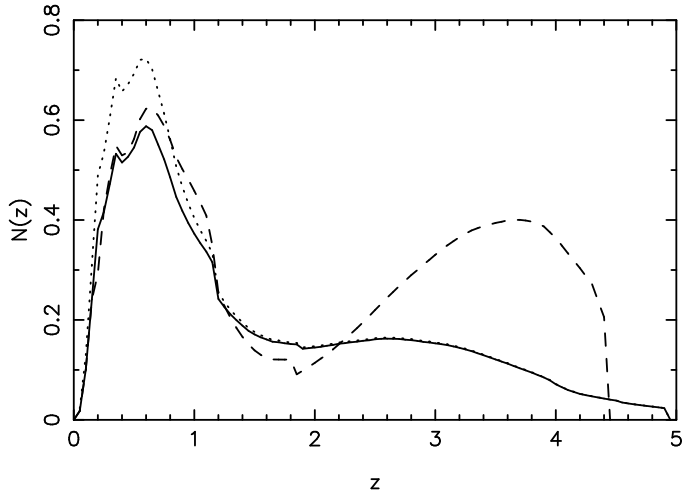}
\caption{Redshift distribution of arclets in A370 for the following 
selection criteria: $R_{675W}<23.5$,
$\mu_R<25.5$ and $a/b>2$. The solid line corresponds to the model $q_0=0.5$ 
and the dashed line is
for $q_0=0$. The dotted line corresponds to the ellipticity distribution
of equation \ref{ell_2} with $q_0=0.5$.} 
\label{redshift_675}
\end{figure}

The next step is to consider the redshift distribution of the
arclets. Figure \ref{redshift_675} shows the predicted one with the
selection criteria close to the observational limits for faint 
object
spectroscopy ($R_{675}<23.5$, $\mu_R<25.5$ and $a/b>2$). The
comparison with real data is not easy at present because no well defined
sample of arclets, complete in magnitude, has yet been spectroscopically
explored. The redshift distributions displayed in Figure
\ref{redshift_675} present a prominent peak at $z\simeq 0.6$ and a
secondary one at $z > 2$ due essentially to the contribution of bright
and young Elliptical-type galaxies in their process of strong star
formation. This second peak is strongly attenuated compared to what
was discussed in B\'ezecourt et al. (1998a) where there was clearly a strong
excess in the high redshift tail. Taking into account the ellipticity
distribution of the sources has fixed this problem. 

In order to see how stable the redshift distribution is with respect to 
the ellipticity distribution, another $p(\tau)$ is considered following
Ebbels (1998): 
\begin{equation}
p(\tau) \propto \tau \, \exp \left[- \left({\tau \over 0.20}\right)^{0.85}
\right]
\label{ell_2}
\end{equation}
representative of objects with $18<I<25.5$ while equation \ref{ell} was 
derived with bright objects ($18<I<22$). As this distribution corresponds
to more elongated objects than in equation \ref{ell}, more arclets appear at
low $z$ (Figure \ref{redshift_675}, dotted line). However, the total 
number of arclets is increased by only 11\%.

Analysis of Figure \ref{redshift_675}
suggests several comments:
 
\begin{itemize}
\item Compared with the redshift distribution of field galaxies within the
same observational limits, the arclet distribution is biased towards
more distant objects. We can take advantage of this
modification of the redshift distribution to select more distant
galaxies in a redshift survey sample. This is particularly true near
the central part of the lens where very high redshift galaxies
($z>2.5$) have already been found (\cite{ebbels96},
\cite{trager97}, \cite{franx97}, \cite{frye98}, 
\cite{pello98a}). 
\item The two galaxy evolution models presented in this paper 
($q_0=0$ and $q_0=0.5$) have been validated in empty fields where the
number counts and redshift distribution of galaxies were compatible
with the observed ones. However they present significant differences
at high redshift for arclets.  The $q_0=0$ model with ``Pure
Luminosity Evolution" (PLE) predicts more high redshift arclets than
the $q_0=0.5$ model where number density evolution is included. This
different behavior of the redshift distribution at $z>2$ is
encouraging as it could be a way to distinguish the two scenarios
by
analyzing the redshift distribution of arclets with $z>2$.  
Although it is presently difficult to speculate on the true
fraction 
of very high redshift objects in well defined samples of arclets, this 
population of very high redshift arclets does not seem to dominate the 
actual samples. Hence we favor
the $q_0=0.5$ model corresponding to a merging scenario of galaxy
formation and evolution.

\item Because the high redshift domain is sensitive 
to many uncertainties included in the evolutionary code (dust
obscuration in the UV, uncertainties in the slope of the IMF for
massive stars or in their UV tracks, influence of the short time scale
phenomena), we must be cautious on the above conclusions.
Furthermore, the galaxy evolution models assume that all the galaxies
form instantaneously at a given redshift, which is a clear limitation
of these models (we have $z_{\rm form} = 4.5$ for $q_0=0$, which gives
for the present day galaxies an age of 16 Gyr, and $z_{\rm form} = 5$ 
for $q_0=0.5$,
corresponding to an age of 12.2 Gyr). A more phenomenological 
model of
galaxy formation that follows in detail the mass evolution of
galaxy
sub units (\cite{baugh98}) may improve this simple description, and
would likely be the best to be compared with the current data.
\end{itemize}

\subsection{Depletion curves in Abell 370}
\begin{figure}
\centerline{
\psfig{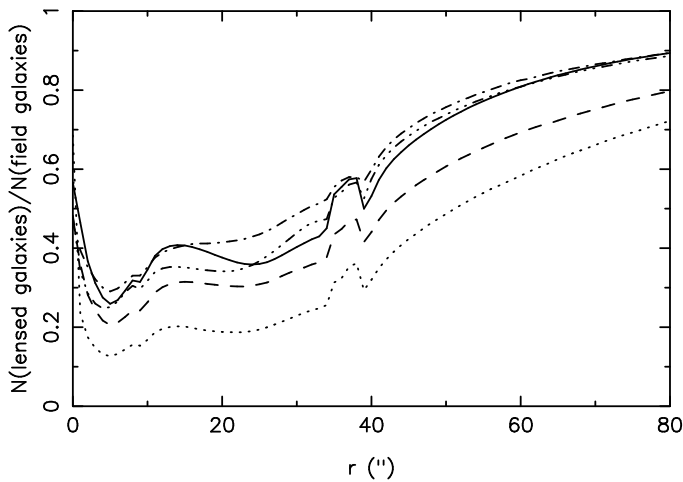}
}
\caption{Depletion of lensed object in different filters and limiting 
magnitudes: $U<23$ ({\bf-- $\cdot$ $\cdot$
$\cdot$ --}), $B<24.5$ ({\bf---}), $R<24$ ({\bf-- -- --}), $I<23$ 
({\bf-- $\cdot$ --}) 
and $K<21$ ({\bf $\cdot$ $\cdot$ $\cdot$}) 
for $q_0=0$. The small excess near $r=40''$ is due to the local
magnification 
enhancement of an individual cluster galaxy.}
\label{depletion}
\end{figure}

For a simple description of the background population, it is easy to show 
that the number density of objects brighter than a given magnitude $m$ 
behind a lens is:
\begin{equation}
 N(<m, A) = N_0 (<m) \, A^{2.5 \alpha - 1} 
\end{equation}
where $N_0 (<m)$ is the number density in blank field, $A$ is the
magnification and $\alpha$ is the logarithmic slope of field number
counts.  An excess or a lack of objects is then expected for a slope
steeper or shallower than 0.4 respectively.  In practice, optical
number counts of field galaxies show a slope smaller than 0.4 in
nearly all wavebands (except in U and B at relatively bright
magnitudes), so a``depletion'' is expected in most cases, being
more pronounced at longer wavelengths (Fort et al. 1996, Taylor et
al. 1998).

With our model, it is possible to compute radial depletion curves
instead of global number counts, for any filter or magnitude
range. A clear illustration of the wavelength dependence of the
magnification bias is given by the ratio of the number of lensed
objects expected in a given area in the field of Abell 370 over the
number of field galaxies in the same area (Figure
\ref{depletion}). One can note that the predicted intensity of the
depletion is higher at longer wavelengths, as expected, because of
the
shallower slope of the field galaxies counts. Moreover, because of the
flattening of the counts at faint magnitudes, the depletion is also
very sensitive to the magnitude threshold (see the I curve in Figure
\ref{depletion}). UV counts have a slope 
larger than 0.4 at bright magnitudes ($U<23$) but it quickly decreases
at fainter levels. This fast flattening is due to the Lyman break that
goes through the red limit of the U filter at redshift $z\simeq
2.5$. Hence, the lack of objects at faint magnitudes in $U$ produces
the depletion curve shown in Figure \ref{depletion} (see Paper II for 
more details of the UV modeling).

The detection of this magnification bias is highly dependant on the
poissonian noise of the background sources and the contamination by
cluster members and foreground objects. This is critical in cluster
cores where the density of objects is very low, of the order of a
few units per arcmin$^2$. Hence, these very poor statistics cannot
bring valuable information on the sources redshift distribution as most of
the information is expected in the dip of the depletion curve.  
Only massive clusters are able
to show such a depletion curve but the recovery of the sources
redshift distribution seems out of reach with present datasets.

\section{Conclusions}
In this paper, we have presented a new analysis of the cluster lens
Abell 370. Thanks to the measurement of the redshift for the B2--B3
gravitational pair ($z=0.806$) and the identification of several new
multiple arclets in the WFPC2 image, a more accurate and well
constrained model of the mass distribution in Abell 370 is proposed, 
including galaxy scale components. We
studied the spectral energy distribution (SED) of arclets B3 and A5
 which are found to be both young, low metallicity
star forming objects without strong interstellar extinction.

The lens model has been used to study the background population of
galaxies. Taking the
galaxy ellipticity distribution into account induces significant
changes in the predicted redshift distribution of arclets.
The excess of very distant sources found
in previous analyses is now strongly attenuated and arclets at low
redshift are recovered.
For the two prescriptions used on galaxy evolution, 
although they both reproduce well the number counts in 
empty fields and in cluster lenses, the ``Pure Luminosity Evolution''
model in a low density universe over predicts the number 
of high redshift ($z>2$) galaxies, compared to what is currently observed. 
If this effect is observationally confirmed it   may constrain the number
density evolution of galaxies in order to interpret the redshift 
distribution of arclets.

To understand further the properties of the background population
through cluster lenses, deep dedicated cluster surveys are needed
in
order to enlarge the number of arclets and have significant
statistical properties. A multi-color approach is favored as it
can
help in constraining the redshift distribution. This is particularly true
for
those unresolved sources for which no shape information is available and 
only the spectral energy distribution can discriminate between background 
objects and cluster members. In addition, detailed study of their
spectral energy distribution will be useful to characterize properties like
the luminosity function, dust extinction and star formation histories of
distant galaxies (e.g. Pell\'o et al. 1998b).

\acknowledgements
We thank R. Pell\'o, R.S. Ellis  and Y. Mellier 
for many fruitful discussions and encouragements. 
This research has been conducted under the auspices of a European 
TMR network programme made possible via generous financial support 
from the European Commission ({\tt http://www.ast.cam.ac.uk/IoA/lensnet/}). 
This work was also supported by the Programme National de Cosmologie and CNRS.

\end{document}